\documentclass[twoside,showpacs,superscriptaddress,twocolumn,floatfix,a4paper,pra]{revtex4-1}

\usepackage{siunitx, amsmath, multirow, hyperref}
\usepackage{graphicx}
\usepackage[utf8x]{inputenc}
\usepackage[T1]{fontenc}
\usepackage[normalem]{ulem}

\usepackage{lmodern}
\usepackage[rgb]{xcolor}
\graphicspath{{fig/}}

\begin{document}

\title{Measuring evolution of a photon in an interferometer with spectrally-resolved modes}

\author{Marek Bula}
\affiliation{RCPTM, Joint Laboratory of Optics of Palacký University and Institute of Physics of Czech Academy of Sciences, 17. listopadu 12, 771 46 Olomouc, Czech Republic}

\author{Karol Bartkiewicz} 
\affiliation{Faculty of Physics, Adam Mickiewicz University, PL-61-614 Pozna\'n, Poland}
\affiliation{RCPTM, Joint Laboratory of Optics of Palacký University and Institute of Physics of Czech Academy of Sciences, 17. listopadu 12, 771 46 Olomouc, Czech Republic}

\author{Antonín Černoch} 
\affiliation{RCPTM, Joint Laboratory of Optics of Palacký University and Institute of Physics of Czech Academy of Sciences, 17. listopadu 12, 771 46 Olomouc, Czech Republic}

\author{Dalibor Javůrek}
\affiliation{RCPTM, Joint Laboratory of Optics of Palacký University and Institute of Physics of Czech Academy of Sciences, 17. listopadu 12, 771 46 Olomouc, Czech Republic}

\author{Karel Lemr} 
\affiliation{RCPTM, Joint Laboratory of Optics of Palacký University and Institute of Physics of Czech Academy of Sciences, 17. listopadu 12, 771 46 Olomouc, Czech Republic}

\author{Václav Michálek}
\affiliation{Institute of Physics of Czech Academy of Sciences, Joint Laboratory of Optics of PU and IP AS CR, 17. listopadu 50A, 772 07 Olomouc, Czech Republic}

\author{Jan Soubusta}
\affiliation{Institute of Physics of Czech Academy of Sciences, Joint Laboratory of Optics of PU and IP AS CR, 17. listopadu 50A, 772 07 Olomouc, Czech Republic}

\date{\today}

\begin{abstract}
In the year 2013, Danan \emph{et al.} published a paper [Phys. Rev. Lett. {\bf 111}, 240402 (2013)] demonstrating a counterintuitive behavior of photons in nested 
Mach-Zehnder interferometers. The authors then proposed an explanation based on the two-state vector formalism. This experiment and the authors' explanation raised a vivid debate within the scientific community. In this paper, we contribute to the ongoing debate by presenting an alternative experimental implementation of the Danan \emph{et al.} scheme.  
We show that no counterintuitive behavior is observed when performing direct spectrally-resolved detection.
\end{abstract}

\pacs{42.50.-p, 42.50.Dv, 42.50.Ex}

\maketitle

\section{Introduction}

\begin{figure}
  \includegraphics[width = 0.8\columnwidth]{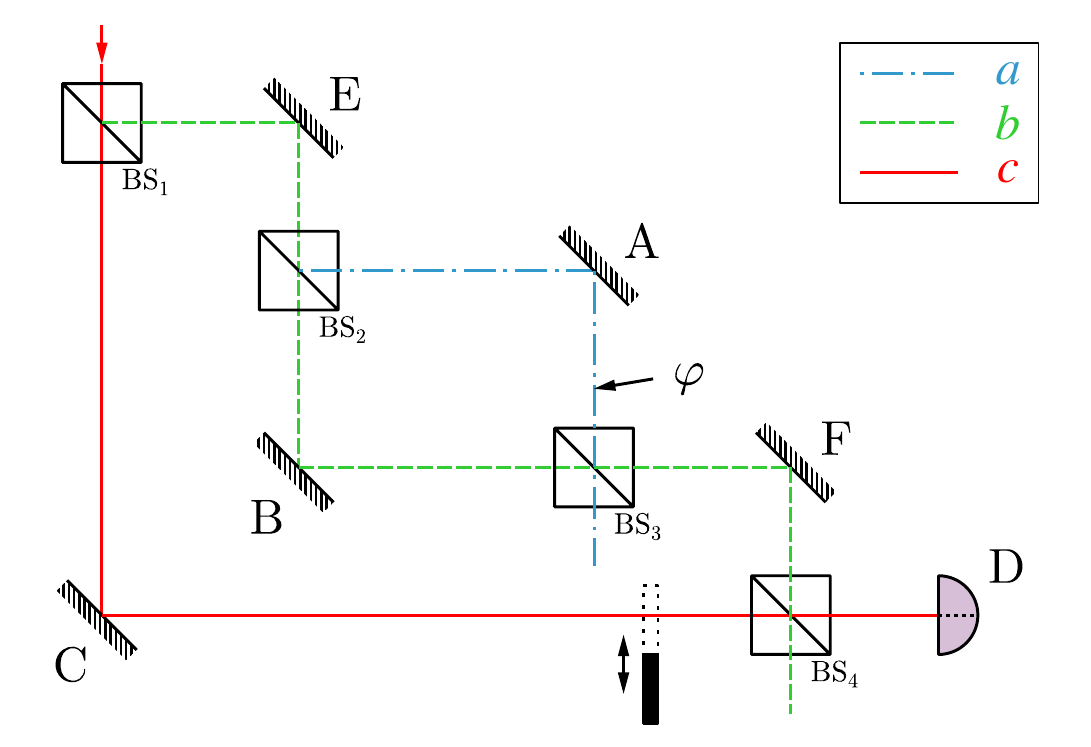}
  \caption{\label{fig_scheme} (color online) Scheme of the original Danan \emph{et al.} experiment 
    \cite{Danan2013}. The counterintuitive result was observed when lower arm of the outer 
    interferometer (path ``c'') was blocked and the inner interferometer was set so that 
    destructive interference occurs in the detected output port. Even the photons known to travel 
    by one specific arm of the inner interferometer (either mirror A, or B) interfered destructively.}
\end{figure}

Over the past three years, Lev Vaidman and his coworkers published several papers discussing a new 
method for analyzing the past of a photon detected at the output of an interferometer
\cite{Vaidman2007,Vaidman2013,Vaidman2014,Danan2013}. 
This method, based on the two-state vector formalism (TSVF), raised a vivid debate within the 
scientific community. In their seminal paper on this topic \cite{Danan2013}, Danan \emph{et al.} 
presented experimental evidence for the validity of this approach. In that experiment, two nested 
Mach-Zehnder interferometers were constructed with mirrors A, B, C, E and F vibrating at 
distinct frequencies (see Fig. \ref{fig_scheme}). Harmonic analysis of the output 
signal allowed to determine whether a photon passed by a specific mirror. Three different 
configurations were tested in the original experiment. In one of these configurations (labeled as 
``c'' in the original paper) a highly counterintuitive outcome was obtained. This result was 
observed when the outer interferometer was disabled by blocking its lower arm (path ``c'') and 
thus only the inner interferometer was operational. Phase shift in this inner interferometer was 
adjusted so that the two indistinguishable photon paths interfered destructively in the detected 
output port ($\varphi = 0$).

Quite surprisingly, even the photons, whose paths in the interferometer had been 
identified by the harmonic analysis, interfered destructively. This effect seemingly violates the 
Englert-Greenberger-Yasin duality relation, which is related with the wave-particle duality of 
photons \cite{Greenberger88,Scully91,Jaeger95,Herzog95,Englert96,Englert99}.
The authors of Ref. \cite{Danan2013} argued that this effect can be explained by means of the TSVF.
This particular experiment and the unusual interpretation of the results were subsequently 
followed by a number of papers \cite{Saldanha2014, Potocek, Salih, BCJLSS, Wu15, LVC}. 
Authors of these papers for instance provided an alternative classical formalism 
\cite{Saldanha2014, Potocek} for the experiment or questioned its specific implementation 
\cite{Salih, BCJLSS}.

In 2015, we theoretically analyzed the Danan \emph{et al.} experiment \cite{Danan2013} and proposed a standard quantum-mechanical description of the action of vibrating mirrors introducing distinguishable modes \cite{BCJLSS}. In that paper, we also argued that the method used to process detected light in the Danan \emph{et al.} experiment is not suitable because it neglects part of the signal. This is then incorrectly interpreted as observing destructive interference. In Ref. \cite{BCJLSS}, we proposed a different measurement technique to reveal the path of the photon inside the interferometer. The suggested detection method is based on direct spectral power density measurement. We predicted identical results in all configurations of the Danan \emph{et al.} experiment except for the counterintuitive case with blocked path ``c''. Our detection method predicts quite intuitive results even in this configuration.

In this paper, we report on an experimental implementation of our measurement method. Instead of the vibrating mirrors we use {different} frequency filters to shape spectra of the photons propagating in {the upper and lower arm of} the interferometer. Thus, we ``leave a mark'' on them which can then be used to extract which-path information during the detection stage. Our detection method and the one presented in the original paper by Danan \emph{et al.} give incompatible results only when the path ``c'' is blocked (see Fig. \ref{fig_scheme}). In this case, the outer interferometer does not influence the results and therefore we can test our detection method using only one Mach--Zehnder interferometer playing the same role as the inner interferometer in the Danan \emph{et al.} experiment. Our experimental setup is sketched in Fig. \ref{fig_setup}.

Our experimental setup in described in Sec. \ref{sec:exper}.  In Sec. \ref{sec:meas}, we describe the performed spectrum measurement and the procedure to reconstruct the photons which-path information in a way analogous to the Danan \emph{et al.} experiment \cite{Danan2013}. In the original work \cite{Danan2013}, temporal evolution of the overall intensity was recorded and post-processed. From the post-processed data, the intensity spectrum was computed. In contrast to that, we measure the intensity spectrum directly and demonstrate that it is in agreement with our theoretical model presented in Ref. \cite{BCJLSS}.
Additionally, we show that the obtained measurements fulfill the Englert-Greenberger-Yasin inequality 
\cite{Greenberger88,Englert96}. Moreover, in Sec.~\ref{sec:simul}, we use a post-processing method similar to the one used in the original experiment \cite{Danan2013}. Based on this approach, we explain the apparent loss of the signal that can be incorrectly attributed to destructive interference.

\section{Experimental setup} 
\label{sec:exper}

\begin{figure}
  \includegraphics[scale=0.3]{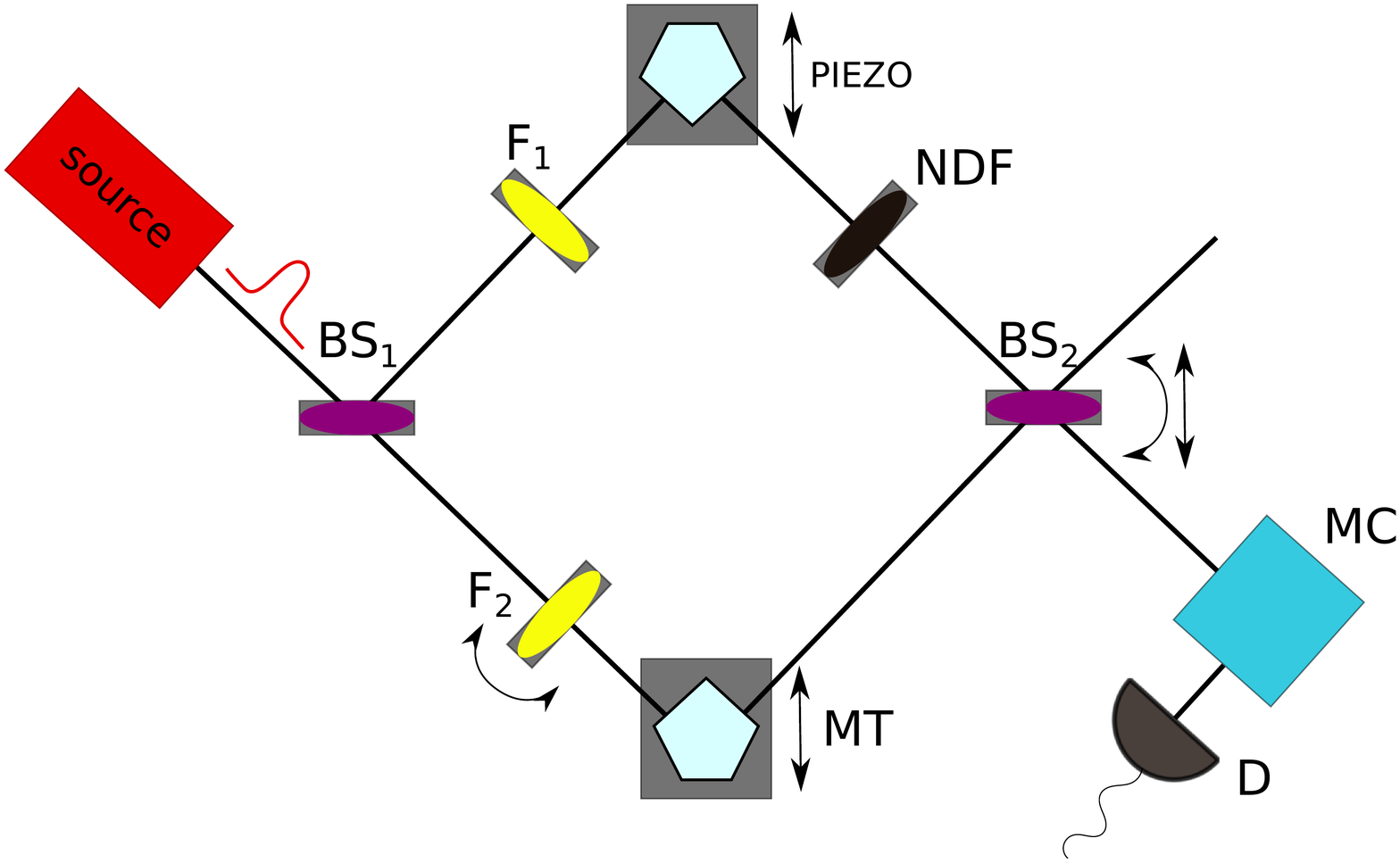}
  \caption{(color online) Experimental setup as described in the text. Individual components are labelled as follows: 
  BS$_{1,2}$ $-$ beam splitters, NDF $-$ neutral-density filter, F$_{1,2}$ $-$ narrow-band filters, 
  MT $-$ motor translation, MC $-$ monochromator, D $-$ detector.
  \label{fig_setup}}
\end{figure}

Our experimental setup is based on a Mach--Zehnder interferometer as depicted 
in Fig.~\ref{fig_setup}. The interferometer consists of two polarization 
independent beam splitters BS$_{1,2}$ and two pentaprisms. The input light beam is 
generated by a modelocked femtosecond laser Mira (Coherent) with central wavelength of 
\SI{826}{\nano\meter}, \SI{10}{\nano\meter} bandwidth and the typical mean power of 
\SI{1}{\watt}. The light is coupled into the 
interferometer through the beam splitter BS$_1$. Motorized translation stage in the lower 
arm is used to balance the lengths of both interferometer arms. The piezo translation 
stage in the upper arm is used to change the relative phase between these arms. 
In this experiment, we used two identical narrow-band spectral filters F$_{1,2}$ with 
transmission bandwidth of \SI{3}{\nano\meter} centered at \SI{826}{\nano\meter}. F$_2$ 
was mounted on a rotation stage. By rotating the filter, we were able to shift its transmission window to shorter wavelengths. In our experiment, filters F$_1$ and F$_2$ introduce distinguishability between the interferometer arms just as the vibrating mirrors did in the original paper \cite{Danan2013}. Finally, the beam splitter BS$_2$ couples the light from both the interferometer arms. Due to technological imperfections, BS$_1$ and BS$_2$ 
have not perfectly balanced splitting ratios. Also the overall transmissivity of F$_2$ depends on its 
rotation. Therefore, a neutral density filter (NDF) was inserted into the upper arm 
to balance effective losses. The Monochromator (MC) Jobin Yvon Triax 320 is located in one of the 
interferometer output ports and provides the capability to discriminate spectral 
components of the beam (shown in Fig.~\ref{fig_setup}). Light was transferred from the output port of BS$_2$ to this monochromator using a single-mode fiber to maximize spatial mode indistinguishability. Detection was performed at the output of the monochromator using a Power Meter 
PM120 by Thorlabs (labeled D).

For the preliminary adjustments of the setup, both filters F$_{1,2}$ were rotated 
perpendicular to the light beam direction and the detector was placed directly to the BS$_2$ 
output port bypassing thus the monochromator. First, we ensured precise coupling 
of the beam to the setup. We checked, that the polarization remains unchanged while being
transmitted or reflected on beam splitters. The second part of the adjustment procedure consists of several steps, which have to be repeated for each setting of the measurement. Balanced output intensities from both arms are achieved by rotating NDF wheel with gradient absorption. Then lengths of the arms were equalized by positioning the translation stage MT. Accurate setting of the MT position was adjusted by finding maximum of the 
autocorrelation function of the signal. We were able to reach visibility of the interferometer 
typically about \SI{98}{\percent} in the initial configuration with filter F$_2$ inserted perpendicularly to the beam. The interferometer was suffciently stable to scan the interference fringes by varying voltage on the piezo-driven translation stage.

\section{Measurement and results}
\label{sec:meas}

The purpose of our experiment is to recreate a similar situation as in the original research paper by Danan \emph{et al.} \cite{Danan2013}. But instead of using vibrating mirrors, we encode the which-path information directly into different shapes of spectra in the upper and lower interferometer arm. Further to that, our experimental configuration allows us to tune the distinguishability between the interferometer arms. This is achieved by rotation of the second filter F$_2$. In subsequent paragraphs, we label three spectral modes A, B and E in accordance to the labelling of mirrors by Danan \emph{et al.} \cite{Danan2013}. In the original experiment drawn in Fig. \ref{fig_scheme}, mirrors A and B were placed in the upper and lower arm of the interferometer respectively. Mirror E stood in front of that interferometer. Observing vibration frequencies A and B thus gave information about the propagation of light in the upper and lower arm respectively. Frequency peak E did not provide any which-path information. Similarly to that, we associate spectral modes A and B with the modes maximizing the predictability of photons propagating in the upper (mode A) and lower (mode B) arm. Mode E was chosen to give least amount of predictability. In the original experiment \cite{Danan2013}, the modes with Gaussian spectral profile reflected from mirrors A and B are probabilistically distinguishable by detection of their transverse spatial modes. The beams are indistinguishable if the mirrors are deviated by equal angle from plane of the interferometer. Therefore, except particular times, the beams reflected from mirrors A and B are probabilistically distinguishable. In our experiment, the same is true for beams with quasi-Gaussian spectra shaped by filters F$_1$ and F$_2$. The probabilistic distinguishability in transversal spatial modes is interchanged with probabilistic distinguishability in spectrum.

We have performed several sets of measurements for assorted rotation angles of F$_2$. 
Each measurement set consists of several scans through the frequency spectrum 
in the range from \SI{815}{\nano\meter} to \SI{835}{\nano\meter} with resolution of \SI{0.2}{\nano\meter}. First, we simply measured spectra separately from the upper and lower arm. 
We used this data to calculate theoretical predictions for the visibility. Next, for each 
wavelength we measured the visibility of interference as a measure of indistinguishability 
between the arms. The visibility is calculated from the minimal $I_\mathrm{min}$ and maximal $I_\mathrm{max}$ power density in an interference fringe at a given wavelength $\lambda$ using the formula

\begin{equation}
  V(\lambda)=
     {I_{\mathrm{max}}(\lambda)-I_{\mathrm{min}}(\lambda)
       \over
      I_{\mathrm{max}}(\lambda)+I_{\mathrm{min}}(\lambda)}.
  \label{eq:visibility}
\end{equation}

\begin{figure}[t]
\includegraphics[scale=1]{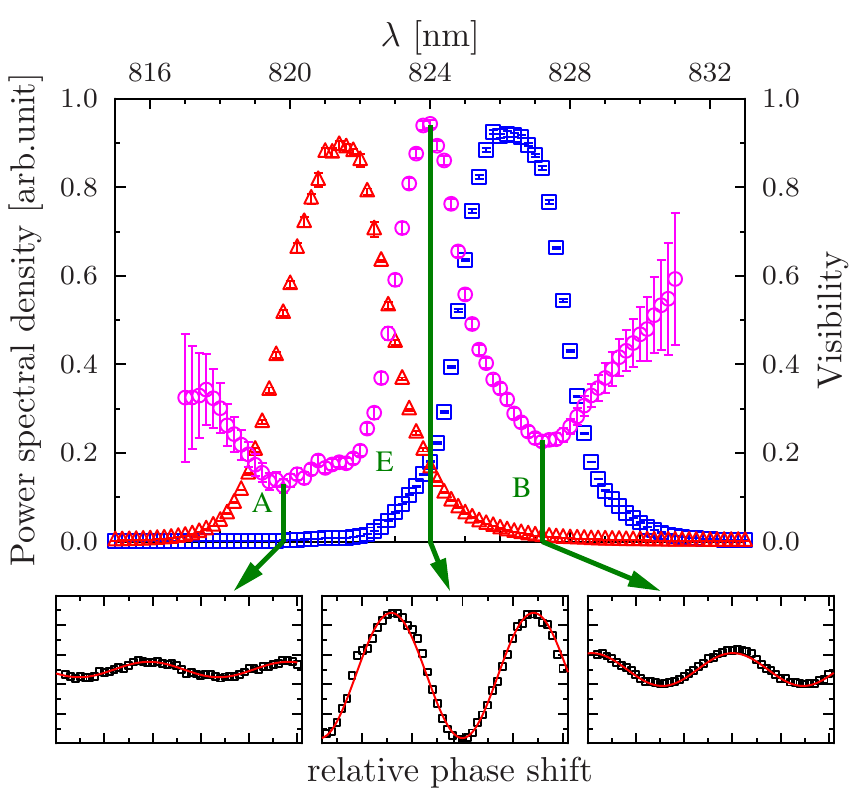} 
\caption{\label{fig_velbloud} (color online) Example of measured data for F$_2$ rotated by 14 deg. resulting in
$\Delta\lambda$ = \SI{4.9}{\nano\meter}. 
Triangles and squares represent independent spectral power density for lower and upper arms. Circles visualize obtained visibility. Vertical lines labelled A, E and B mark the selected wavelengths corresponding to maximum and two minima of the visibility. Interference fringes (spectral power density as function of mutual phase shift between interferometer arms) are visualized for the selected modes A, B and E.
}	
\end{figure}

The uncertainty of the power density measurement was estimated from the typical fluctuation 
of the power measurement of the power meter D. Typical example of one set of these spectrum scans is depicted in Fig.~\ref{fig_velbloud} including interference fringes at three selected wavelengths. 

Based on this spectrum scan, we have selected three wavelengths corresponding to the maximum and two minima of visibility. These wavelengths are labeled E for maximum of the visibility and A, B for the two minima. Both arms contribute by equal power density at the wavelength E, thus we observe maximal interference visibility at this wavelength for the given rotation of filter F$_2$. On the other hand, frequencies A and B are chosen to maximize the distinguishability of the two respective photon paths in the interferometer.

\begin{figure}
\includegraphics[width=0.48\textwidth]{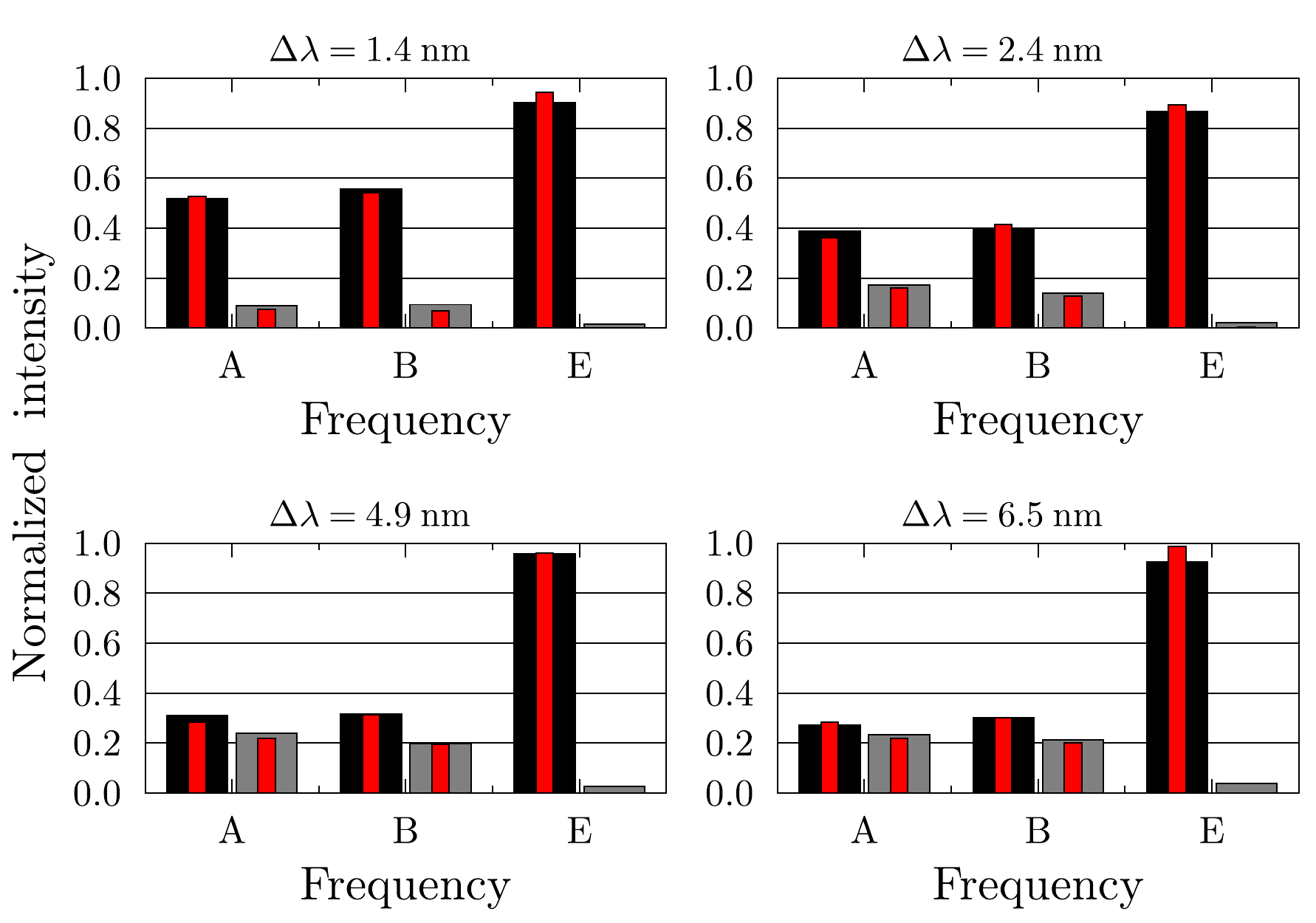}

\caption{(color online) Processed measurement data for four rotations of F$_2$, which correspond to spectral distance $\Delta\lambda = 1.4$, 2.4, 4.9 and 6.5 nm. 
Black bars represent normalized maximum spectral power densities (constructive interference) for wavelengths A, B, and E. 
Grey bars, in a similar way, represent spectral power density minima (destructive interference). Theoretical predictions 
are shown using the inner red bars.\label{fig_triVelbloudi}}
\end{figure}

\begin{table*}
\caption{Measured minimum and maximum normalized power densities in an interference fringe for various rotations of the filter~F$_2$. Theoretical prediction shown below is based on individual spectra taken from the interferometer arms separately.}
\centering
\footnotesize
\begin{ruledtabular}
 \begin{tabular}{cccccccc}
  & & \multicolumn{2}{c}{\bf Frequency mode A} & \multicolumn{2}{c}{\bf Frequency mode B} 
    & \multicolumn{2}{c}{\bf Frequency mode E} \\
  $\Delta\lambda$ & & $I_\mathrm{max}$ & $I_\mathrm{min}$ & $I_\mathrm{max}$ & $I_\mathrm{min}$ 
    & $I_\mathrm{max}$ & $I_\mathrm{min}$ \\
\\
  \multirow{2}{*}{1.4 nm}
    & experiment & $0.519\pm 0.005$ &$0.089 \pm 0.002$ & $0.555 \pm 0.007$ & $0.094 \pm 0.003$ 
      & $0.902 \pm 0.006$ & $0.014 \pm 0.001$ \\ 
    & theory & $0.525$ & $0.076$ & $0.542$ &$0.067$&$0.945$&$0.001$ \\
  \multirow{2}{*}{2.4 nm}
    & experiment & $0.388 \pm 0.006$ & $0.171 \pm 0.004$ & $0.398 \pm 0.005$ & $0.140 \pm 0.003$ 
      & $ 0.868 \pm 0.006$ & $ 0.020 \pm 0.001$ \\
    & theory & $ 0.360$ & $0.160$ &$0.414$ & $0.127$ & $0.894$ & $0.003$ \\
  \multirow{2}{*}{4.9 nm}
    & experiment & $0.309 \pm 0.007$ & $0.240 \pm 0.006$ & $0.318 \pm 0.005$ & $0.199 \pm 0.003$ 
      & $0.958 \pm 0.020$ & $0.028 \pm 0.005$\\
    & theory & $0.284$ & $0.218$ & $0.312$ & $0.195$ & $0.962$ & $0.0004$\\
  \multirow{2}{*}{6.5 nm}
    & experiment & $0.274 \pm 0.005$ & $0.234 \pm 0.005$ & $0.300 \pm 0.006$ & $0.214 \pm 0.005$ 
      & $0.925 \pm 0.030$ & $0.039 \pm 0.010$ \\ 
    & theory & $0.274$ & $0.219$ & $ 0.303$ & $0.202$ & $0.987$ & $0.00004$
\label{tab_measurement}
\end{tabular}
\end{ruledtabular}
\vline
\end{table*}

Four representative results of the normalized power densities are summarized in Tab.~\ref{tab_measurement} and  visualized in Fig.~\ref{fig_triVelbloudi}.
Black bars depict power density maxima ${I_{\rm max}}$ corresponding to constructive interference, and grey bars depict power density minima ${I_{\rm min}}$ corresponding to destructive interference. To obtain normalized values of ${I_{\rm max}}$ and ${I_{\rm min}}$, we have divided the measured spectral power densities by four times the spectral power density measured separately with one arm blocked. The factor 4 arises from the fact that only one quarter of the signal entering the interferometer leaves by the detected output port if one arm is blocked. As it is evident from the experimental setup shown in Fig.~\ref{fig_setup}, the signal is divided in half on each of the two balanced beam splitters. Inner red bars depict theoretical predictions calculated using the formulas for perfect constructive and destructive interference
\begin{eqnarray}
{I_{\rm max}} &=& \left(\sqrt{{I_1}}+\sqrt{{I_2}}\right)^2,\nonumber\\
{I_{\rm min}} &=& \left(\sqrt{{I_1}}-\sqrt{{I_2}}\right)^2,
\end{eqnarray}
where ${I_1}$ and ${I_2}$ stand for normalized spectral power densities measured with
the other interferometer arm blocked. Experimental data are in good agreement with these theoretical predictions.

As a next step, we have subjected the obtained data to the Englert-Greenberger-Yasin inequality test
\cite{Greenberger88,Englert96} expressed in the form of
\begin{equation}
\label{eq:EGY}
V^2(\lambda) + D^2(\lambda) \leq 1,
\end{equation}
where $V(\lambda)$ stands for visiblity and $D(\lambda)$ for distinguishability of the paths given by absolute value of the difference between normalized power densities from the first \textbf{($I_1$)} and from the second arm \textbf{($I_2$)}
\begin{equation}
\label{eq:D}
D(\lambda) =  \left| \frac{I_1(\lambda) - I_2(\lambda)}{I_1(\lambda) + I_2(\lambda)}\right|.
\end{equation}
By using this test, we can verify the consistency of our measurements with standard quantum-mechanical limits. Our results are summarized in Tab.~\ref{tab_VD}. Note, that in the case of the original experiment by Danan \emph{et al.}, the inequality (\ref{eq:EGY}) is seemingly violated, because the frequency modes A and B allegedly interfere with high visibility despite the fact, that they provide exact which-path information ($D\to 1$).

Our obtained results show that, quite intuitively, the visibility decreases as the frequency modes A and B become more distinguishable (higher rotation angles of filter F$_2$). For perfectly distinguishable modes, visibility should drop to zero. As a result, the signal at wavelengths corresponding to frequency modes A and B is nearly constant independent of the setting of the mutual phase shift between the interferometer arms. 

We observe almost perfect distinguishability when rotating the filter F$_2$ to achieve spectral distance $\Delta\lambda = 6.5$~nm (more then twice full width at half maximum of the relevant peaks). In this case, the situation is analogous to the experiment by Danan \emph{et al.} with the outer interferometer blocked (see Fig.~\ref{fig_scheme}). Even when setting destructive interference to occur in the detected output port, the probability of observing photons at frequencies either A or B are nonzero [see grey bars in Fig.~\ref{fig_triVelbloudi} ($\Delta\lambda = 6.5$ nm)]. On the other hand, the probability of observing photons with frequency E vanishes. Measured probabilities of observing photons at frequencies A, B and E are reaching values 0.234$\pm0.005$, 0.214$\pm0.005$ and 0.039$\pm0.010$, respectively. Our theoretical model predicts the values to be 0.25, 0.25 and 0 respectively \cite{BCJLSS}. We attribute these small discrepancies to non-perfect distinguishability of the two modes A and B and imperfect interference at frequency E.
\begin{table*}
\caption{Measured values of visibility $V$ and distinguishability $D$ for various setting of the filter F$_2$ rotation.}
\centering
\footnotesize
\begin{ruledtabular}
 \begin{tabular}{ccccccccccc}
  & \multicolumn{3}{c}{\bf Frequency mode A} & \multicolumn{3}{c}{\bf Frequency mode B} 
    & \multicolumn{3}{c}{\bf Frequency mode E} \\
  $\Delta\lambda$ & $V$ & $D$ & $V^2 + D^2$ & $V$ & $D$ & $V^2 + D^2$ & $V$ & $D$ & $V^2 + D^2$ \\
  \\

1.4 nm & $0.71\pm 0.01$ & $0.66\pm 0.01$ & $0.94\pm 0.02$
& $0.71\pm 0.01$ & $0.64\pm 0.02$ & $0.91\pm 0.02$
& $0.97\pm 0.01$ & $0.06\pm 0.01$ & $0.94\pm 0.01$\\

2.4 nm & $0.39\pm 0.01$ & $0.92\pm 0.02$ & $1.00\pm 0.04$
& $0.48\pm 0.01$ & $0.85\pm 0.02$ & $0.95\pm 0.03$
& $0.95\pm 0.01$ & $0.11\pm 0.01$ & $0.93\pm 0.01$\\


4.9 nm & $0.13\pm 0.01$ & $0.99\pm 0.03$ & $1.00\pm 0.05$
& $0.23\pm 0.01$ & $0.97\pm 0.02$ & $1.00\pm 0.04$
& $0.94\pm 0.01$ & $0.04\pm 0.01$ & $0.89\pm 0.01$\\

6.5 nm & $0.08\pm 0.01$ & $0.99\pm 0.02$ & $0.99\pm 0.05$
& $0.17\pm 0.01$ & $0.98\pm 0.02$ & $0.99\pm 0.05$
& $0.92\pm 0.01$ & $0.01\pm 0.02$ & $0.85\pm 0.01$\\

\label{tab_VD}
\end{tabular}
\end{ruledtabular}
\vline
\end{table*}

\section{Harmonic analysis method}
\label{sec:simul}

Let us now focus on the differences and similarities between our experiment and the measurements 
performed by Danan {\it et al.} in Ref. \cite{Danan2013}. They used vibrating mirrors to deflect 
the beam which left a weak mark on its direction. This weak deviation was then inspected using a quad-cell detector
measuring difference of the photocurrents generated in the both halves of the detector. In our experiment, we used tunable spectral narrow-band filters to leave a which-path mark on the 
propagating photons instead.

\begin{figure}
  \includegraphics[width=0.45\textwidth]{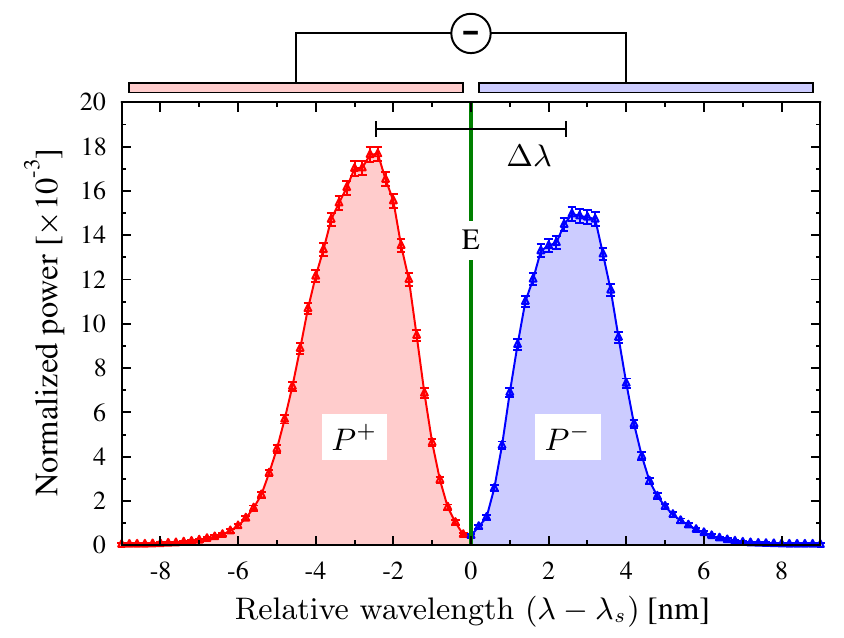}
  \caption{(color online) Spectral dependency of normalized power of minima of interference fringe for $\Delta \lambda = 4.9$ nm. Schematically are depicted two parts of detector centered at frequency $\lambda_s$ (spectral mode E). $P^+$ and $P^-$ denote normalized intensities in their respective bins.
  \label{fig_min}}
\end{figure}

For every selected wavelength, we measured an interference fringe. The maximal and minimal power intensities 
($I_{\rm max} (\lambda)$ and $I_{\rm min} (\lambda)$) were used to calculate the visibility. 
The minima of the normalized power interference fringes are plotted in Fig.~\ref{fig_min}. The interferometer was set for destructive interference ($\varphi=0$).

\begin{figure}
  \includegraphics[width=0.45\textwidth]{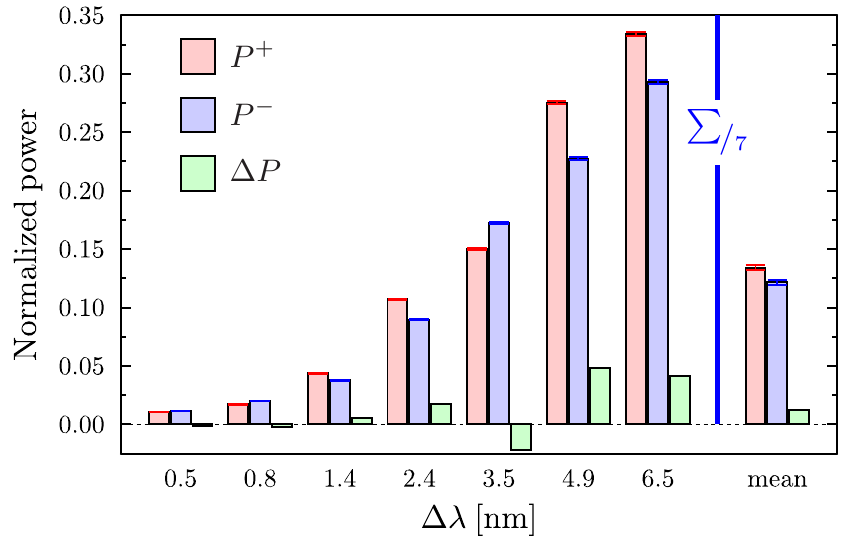} 
  \caption{(color online) The dependence of normalized powers on $\Delta \lambda$, i.e., 
  the spectral distance between transmission maxima of filters F$_1$ and F$_2$. The last set of bars depicts the mean value corresponding to accumulation of the signal across the entire sweep.
  \label{fig_min2}}
\end{figure}

Here the monochromator is formally equivalent to the quad-cell detector distinguishing the above-mentioned marks. While the quad-cell detector registers intensity fluctuations of the resulting interference pattern due to the beam sweeping across its halves, the monochromator distinguishes the spectral marks directly. We emulate the sweep by accumulating the spectra for various rotations of filter F$_2$. Similar to the halves of the quad-cell detector, we have divided the obtained spectra into two bins (see Fig.~\ref{fig_min}). This division between the bins was made at the wavelength $\lambda_s$ associated with the mode E. This mode corresponds to maximum indistinguishability like the central axis of the quad-cell detector. In Fig.~\ref{fig_min2} we summarize normalized powers integrated over the two selected bins
$$ 
 P^+= \frac{\int_{0}^{\lambda_s} I_{\rm min} (\lambda) \, {\rm d} \lambda}
           {\int_{0}^{+\infty}   I_{\rm max} (\lambda) \, {\rm d} \lambda}, 
  \qquad
 P^-= \frac{\int_{\lambda_s}^{+\infty} I_{\rm min} (\lambda) \, {\rm d} \lambda}
           {\int_{0}^{+\infty} I_{\rm max} (\lambda) \, {\rm d} \lambda}, 
$$
and their difference $\Delta P = P^+-P^-$.
 
For large spectral overlap of the filters F$_1$ and F$_2$, the interference fringe 
is deeply modulated and the integrated minimal intensities $P^+$ and $P^-$ are small. 
Increasing the spectral distance of the filters $\Delta\lambda$, the interference 
visibility decreases leading to higher values of binned intensities $P^+$ and $P^-$. Simultaneously the predictability of the photon path increases. But even though there is clearly a distinguishable signal from the two arms of the interferometer, $\Delta P$ simply ignores the which-path information due to the subtraction of $P^+$ and $P^-$. It does not act as a reliable which-path witness. In our experiment, the mean value of $\Delta P = 0.020\pm 0.03$ is about one order of magnitude smaller than mean values of $P^+ = 0.134\pm 0.002$ and $P^- = 0.122\pm 0.002$. Ideally $\Delta P$ should be zero. This is mathematically similar to the Danan {\it et al.} 
experiment, where they measured difference of the photocurrents generated in the two 
halves of the quad-cell detector. Due to the symmetry of the light profile on the 
detector the which-path information was lost.

\section{Conclusions}

In this paper, we have demonstrated an alternative which-path detection method to the original harmonic analysis used in Ref. \cite{Danan2013}. Our method is based on direct frequency resolution of the signal at the output of the interferometer. In contrast to the harmonic analysis, our method yields quite intuitive results that do not violate the Englert-Greenberger-Yasin inequality. We have shown that when the frequency modes corresponding to respective arms of the interferometer are completely distinguishable, the interference vanishes and signal is observed independently on mutual phase shift between the interferometer arms. On the other hand, frequency modes that give no which-path information manifest high interference visibility as expected.

We have also implemented detection method analogical to the one used in the Danan \emph{et al.} experiment \cite{Danan2013}. Our analysis demonstrates how the which-path information gets lost in this process. This supports our original finding \cite{BCJLSS}, that in the Danan \emph{et al.} experiment \cite{Danan2013}, the missing signal at frequencies A and B was caused by unsuitable detection method rather then by an interference effect. We hope that these results would further contribute to the ongoing scientific debate which follows the exciting experiment by Danan \emph{et al.} from 2013.

\section*{Acknowledgements}
K.~B. and K.~L. acknowledge support by the Czech Science Foundation (Grant No. 16-10042Y) and
the Polish National Science Centre (Grant No. DEC-2013/11/D/ST2/02638). V.~M. acknowledges support by the Czech Science Foundation (Grant No. P205/12/0382). D.J. acknowledges support by the Czech Science Foundation (Grant No.~15-08971S). Finally, the authors acknowledge the project No. LO1305 of the Ministry of Education, Youth and Sports of the Czech Republic.


\begin{thebibliography}{99}

\bibitem{Vaidman2007}
L.~Vaidman,
\emph{Impossibility of the Counterfactual Computation for All Possible Outcomes},
Phys. Rev. Lett. {\bf 98}, 160403 (2007).

\bibitem{Vaidman2013}
L.~Vaidman,
\emph{Past of a quantum particle},
Phys. Rev. A {\bf 87}, 052104 (2013).

\bibitem{Vaidman2014}
L.~Vaidman,
\emph{Tracing the past of a quantum particle},
Phys. Rev. A {\bf 89}, 024102 (2014).

\bibitem{Danan2013}
A.~Danan, D.~Farfurnik, S.~Bar-Ad, and L.~Vaidman,
\emph{Asking Photons Where They Have Been},
Phys. Rev. Lett. {\bf 111}, 240402 (2013).

\bibitem{Greenberger88}
D. M. Greenberger and A. Yasin, 
\emph{Stimultaneous wave and particle knowledge in a neutron interferometer},
Phys. Lett. A {\bf 128}, 391 (1988).

\bibitem{Scully91}
M. O. Scully, B.-G. Englert, and H. Walther,
\emph{Quantum optical tests of complementarity},
Nature {\bf 351}, 111 (1991).

\bibitem{Jaeger95}
G. Jaeger, A. Shimony, and L.~Vaidman,  
\emph{Two interferometric complementaries},
Phys. Rev. A {\bf 51}, 54 (1995).

\bibitem{Herzog95}
T. J. Herzog, P. G. Kwiat, H. Weinfurter, A. Zeilinger,  
\emph{Complementarity and the Quantum Eraser},
Phys. Rev. Lett. {\bf 75}, 3034 (1995).

\bibitem{Englert96}
B.-G. Englert, 
\emph{Fringe Visibility and Which-Way Information: An Inequality},
Phys. Rev. Lett. {\bf 77}, 2154 (1996).

\bibitem{Englert99}
P. D. D. Schwindt, P. G. Kwiat, B.-G. Englert, 
\emph{Quantitative wave-particle duality and nonerasing quantum erasure},
Phys. Rev. A {\bf 60}, 4285 (1999).

\bibitem{Saldanha2014}
P.~L.~Saldanha,
\emph{Interpreting a nested Mach-Zehnder interferometer with classical optics},
Phys. Rev. A {\bf 89}, 033825 (2014).

\bibitem{Potocek}
V.~Potoček and G.~Ferenczi,
\emph{Which-way information in a nested Mach-Zehnder interferometer},
Phys. Rev. A {\bf 92}, 023829 (2015).

\bibitem{Salih}
H. Salih,
\emph{Commentary: `Asking photons where they have been' - without telling them what to say},
Front. Phys. {\bf 3}, 47 (2015).

\bibitem{Wu15}
Z.-Q. Wu, H. Cao, J.-H. Huang, L.-Y. Hu, X.-X. Xu, H.-L. Zhang, and S.-Y. Zhu, 
\emph{Tracing the trajectory of photons through Fourier spectrum,}
Opt. Exp. {\bf 23}, 10032 (2015).
 
\bibitem{BCJLSS}
K.~Bartkiewicz, A.~Černoch, D.~Javůrek, K.~Lemr, J.~Soubusta, and J.~Svozilík,
\emph{One-state vector formalism for the evolution of a quantum state through nested Mach-Zehnder interferometers},
Phys. Rev. A {\bf 91}, 012103 (2015).

\bibitem{LVC}
L.~Vaidman,
\emph{Comment on `One-state vector formalism for the evolution of a quantum state
through nested Mach-Zehnder interferometers'},
Phys. Rev. A {\bf 93}, 036103 (2016).

\bibitem{Reply}
K.~Bartkiewicz, A.~Černoch, D.~Javůrek, K.~Lemr, J.~Soubusta, and J.~Svozilík,
\emph{Reply to ``Comment on `One-state vector formalism for the evolution of a quantum state through nested Mach-Zehnder interferometers' ''},
Phys. Rev. A {\bf 93}, 036104 (2016).

\end{thebibliography}

\end{document}